\documentclass[12pt,preprint]{emulateapj}

\slugcomment{Accepted for publication in Apj}

\shorttitle{Integral observation of the PSR B1509-58 PWN}
\shortauthors{Forot et al.}

\begin{document}

\title{High-energy particles in the wind nebula of pulsar B1509-58 as seen by INTEGRAL}
\author{M. Forot\altaffilmark{1,2}, W. Hermsen\altaffilmark{3,6}, M. Renaud\altaffilmark{1,4}, 
P. Laurent\altaffilmark{1,4}, I. Grenier\altaffilmark{1,2}, P.
Goret\altaffilmark{1,2}, B. Khelifi\altaffilmark{5}, and L.
Kuiper\altaffilmark{3}} \altaffiltext{1}{Service d'Astrophysique,
CEA Saclay, 91191, GIF sur YVETTE, France}
\altaffiltext{2}{AIM-UMR 7158, CEA Saclay, 91191 Gif sur yvette,
France} \altaffiltext{3}{SRON Netherlands Institute for Space
Research, Sorbonnelaan 2, 3584 CA, Utrecht, Netherlands}
\altaffiltext{4}{APC-UMR 7164, 11 place M.Berthelot, 75231 Paris,
France} \altaffiltext{5}{Laboratoire Leprince-Ringuet, IN2P3/CNRS,
Ecole Polytechnique, F-91128 Palaiseau Cedex, France} \altaffiltext{6}{
Astronomical Institute Anton Pannekoek, University of Amsterdam, Kruislaan 403,
1098 SJ Amsterdam, Netherlands}

\email{mforot@cea.fr}

\begin{abstract}

We present observations with the INTEGRAL/IBIS telescope of the
wind nebula powered by the young pulsar B1509-58 and we discuss
the spatial and spectral properties of the unpulsed
emission in the $20-200$ keV energy band. The source extension and
orientation along the northwest-southeast axis corresponds to the
jet emission seen at keV and TeV energies. The hard X-ray
spectrum is consistent with the earlier Beppo-SAX measurements. It
follows a power law with a photon index $\alpha = -2.12 \pm 0.05$
up to $160$ keV. A possible break at this energy is found at the
$2.9\sigma$ confidence level. The 0.1-100 keV data are
consistent with synchrotron aging of pairs in the jet and yield a
magnetic field strength of 22-33 $\mu$G for a bulk velocity of
0.3-0.5c. The synchrotron cut-off energy thus corresponds to a
maximum electron energy of 400-730 TeV.

\end{abstract}

\keywords{acceleration of particles -- gamma rays: observations -- shock waves -- pulsars: individual(PSR B1509)}

\section{Introduction}

The young radio pulsar B1509-58 is associated with the supernova
remnant MSH 15-5{\it 2} (G320.4-1.2)
\citep{Seward1982,Manchester1982}. Its spin parameters (a period
of $150$ ms and period derivative of $1.5 \times 10^{-12}$ s
s$^{-1}$) make it one of the youngest and most energetic pulsars
known, with a characteristic age $\tau \sim 1700$ yr and a
spin-down power of $1.8 \times 10^{37}$ ergs s$^{-1}$. It presents
one of the largest magnetic field strength ($1.5 \times 10^{13}$ G)
\citep{Kaspi1994} recorded for an isolated pulsar. Recent X-ray and
$\gamma$-ray images have revealed a very complex pulsar
environment where one can study several manifestations of pulsar
wind nebulae: the shocked wind in an equatorial flow, a powerful
jet, and the interaction and confinement of these wind features
within the surrounding remnant.

ROSAT first detected the wind nebula in X rays, showing its
17'-long tail extending to the southeast along a faint ridge of
radio emission \citep{Greiveldinger1993,Trussoni1996}, later
confirmed by BeppoSAX. The tail length decreases from 17' to 8.5' (at 10
$\%$ level of the central nebula intensity) between 1.6 keV and
10.0 keV \citep{Mineo2001}. High-resolution Chandra images revealed
further details of the non-thermal structure \citep{Gaensler2002};
first, the axis of the elongated nebula (with a position angle of
$150^{\circ} \pm 5^{\circ}$ north through east) which may
correspond to the pulsar spin axis; then two northern toroidal
arcs of emission, 17" and 30" away from the pulsar, that may
correspond to wisps in an ion-loaded equatorial flow as in the
Crab nebula; and a 4'-long bright collimated feature along the
axis which is interpreted as a jet with a velocity $v_{jet} >
0.2c$ \citep{Gaensler2002,Tamura1996,Brazier1997}. The apparent
absence of a counterjet sets a lower limit of 5 to the brightness
ratio between the two sides (if two jets are powered). Doppler
boosting can account for this contrast provided that the jet bulk
velocity and its inclination to the line of sight verify
$\beta_{jet} cos{\theta_{jet}} \sim 0.28$. The wind
carries away at least $0.05\%$ of the pulsar spin down power.

The emission found by H.E.S.S between $280$ GeV and $40$ TeV presents
a comparable morphology \citep{Hess2005}. The elongated nebula
extends on both sides of the pulsar along the X-ray axis, over
6' (1 $\sigma$ level) to the northwest and southeast direction. The $\gamma$-ray
spectrum is well fitted by an $E^{-2.27 \pm 0.23 \pm 0.20}$ power
law and can be explained by TeV electrons up-scattering the
cosmological microwave background and the ambient interstellar
radiation field. Fitting both the TeV emission and the BeppoSAX one
gives a mean magnetic field of $17$ $\mu$G which is in agreement
 with lower limits found for the jet in the equipartition assumption \citep{Delaney2006}.
 At a distance of $5.2 \pm 1.4$
kpc \citep{Gaensler1999}, the bright part of the X-ray
jet, the longer southeast X-ray tail, and the coincident TeV one
extend from 5 to 30 pc on the sky.

We present observations of the PSR B1509-58 region obtained with
the INTEGRAL/IBIS telescope operating in the hard X-ray range,
describing the morphology and spectral properties of the emission
and comparing them with the X-ray and TeV data.

\section{Observation and data analysis}

PSR B1509-58 was observed with the IBIS coded aperture telescope
onboard the INTEGRAL spacecraft. It consists of a dual detection
layer operating between $\sim15~\hbox{keV}$ and $10~\hbox{MeV}$,
and of a tungstene coded mask located 3.2m above the detector. The
first and pixelized detector plane, ISGRI, is composed of $128$ by
$128$ Cadmium--Telluride (CdTe) semiconductor detectors covering
the energy range from $\sim 15~\hbox{keV}$ to $1~\hbox{MeV}$
\citep{Lebrun2003}. Sky images are obtained by deconvolution of the
ISGRI image by the mask pattern. The resulting angular resolution
is $5.7'$ ($1 \sigma$).

The exposure time of about $1.3$ Ms corresponds to dedicated
observations centered on PSR B1509-58 in 2005, and of Galactic Plane
Scan (GPS) observations from $2003$ to $2005$. It
consists of more than 700 science windows of about 30 mn each. The
analysis was done with the standard INTEGRAL OSA software
(version 5.1).

\subsection{Phase resolved analysis}

In order to study the pulsar wind from PSR B1509-58 and to reduce
contamination by pulsed emission from within the light cylinder,
we have analyzed the data from the off-pulse phase interval. To
extract the light curve, the photon UTC arrival times were
converted into the solar system barycentre using the pulsar
position. Photons were then folded into $20$ phase bins using the
ATNF ephemerides
\footnote{$http://www.atnf.csiro.au/research/pulsar/archive$}. In
each phase bin sky images were
obtained from standard deconvolution of ISGRI detector maps. This method provides an
automatically background subtraction and the unpulsed emission is thus clearly visible.

\begin{figure}[!h]
\epsscale{1.2} \plotone{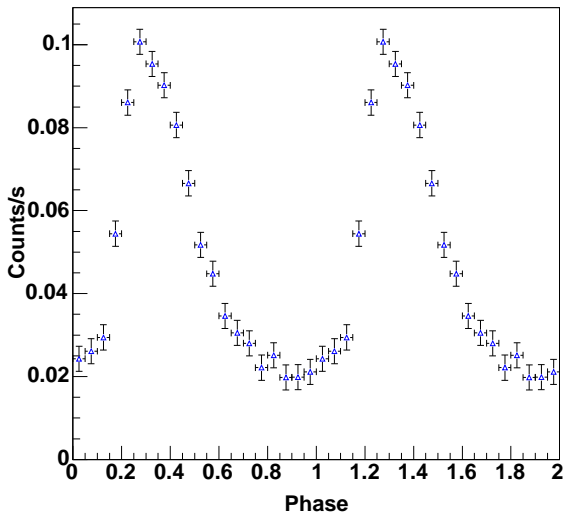}
\caption{The B1509-58 pulsar light-curve between 20 and 50 keV.}
\label{Cdl}
\end{figure}

The resulting light curve is presented in Figure \ref{Cdl}. A
significant unpulsed component is clearly seen in the $0.77-1.07$
phase interval similar to the ones found by BeppoSAX \citep{Cusumano2001} that will be used for the
following imaging and spectral analyzes. A detailed study of the
pulsed emission is deferred to another paper (Hermsen et al., in
preparation).

\subsection{Imaging extended sources with the IBIS/ISGRI telescope}

As shown in Figure \ref{MosaMSH}, the source detected with
IBIS/ISGRI in the off-pulse emission (17-40 keV) appears to be
slightly extended. In the standard OSA software, the flux of a
point-source is given by the peak height of the associated
Point Spread Function \citep{Gros2003}. This is, however, incorrect
for an extended source because of the flux dilution over a region
larger than the PSF width. A method
has been developed for extracting the flux and its associated
error of an extended source seen by IBIS and more generally by
any coded mask telescope using the MURA pattern \citep{Renaud2006}.
This method converts standard flux images into intensity images
(ie flux per sky pixel or per steradian). The total flux is
estimated by summing intensities over the whole source extent.
The dilution factor will be defined as the ratio between the real
flux and the one given by the standard INTEGRAL software. This
factor depends on the source size.

\section{Results}

\subsection{Imaging : an extended source around PSR B1509-58}

\begin{figure}[!h]
\epsscale{1.25} \plotone{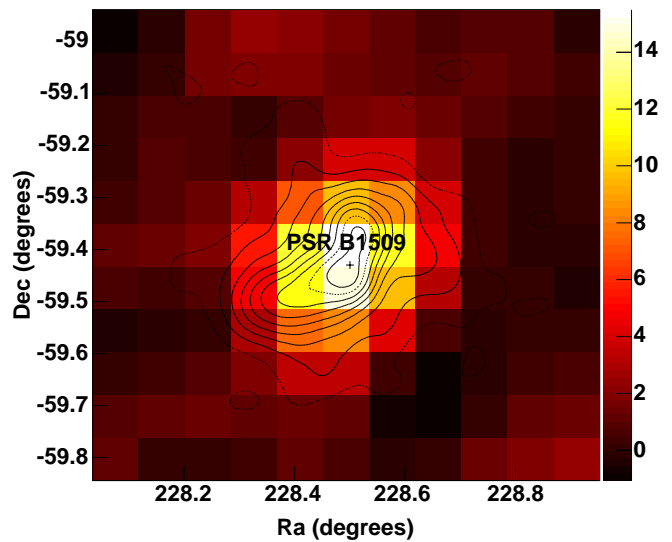}
\caption{Significance
image of the off-pulse emission in the 17-40 keV band around PSR
B1509-58. The coordinates are in J2000. The black
contours outline the TeV emission as seen by H.E.S.S
\citep{Hess2005}.\label{MosaMSH}}
\end{figure}

\begin{figure}[!h]
\epsscale{1.2} \plotone{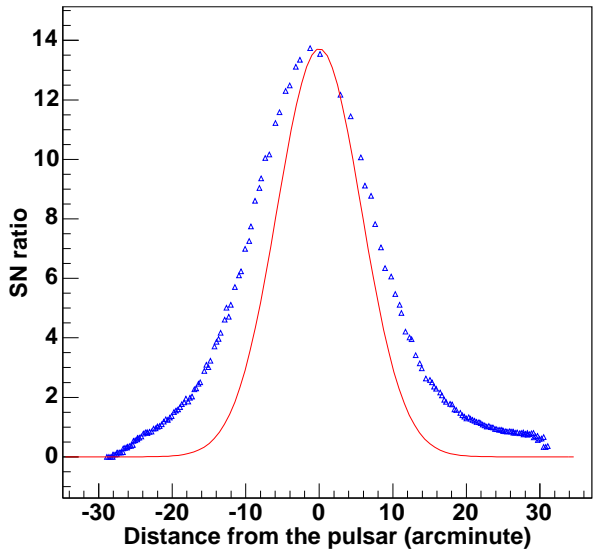}
\caption{ Smoothed profile of the significance map along the NW-SE axis between 17 keV and 40 keV.
The red curve corresponds to the expected profile in a point source case.\label{ProfileSNmap}}
\end{figure}

Maps of the off-pulse emission have been constructed in two energy
ranges, at $17-40$ and $40-100$ keV. The significance map in the
low-energy band is presented in Figure \ref{MosaMSH}. A bright
source is clearly visible at the position of PSR B1509-58. It is
centered on RA$=228.50^{\circ} \pm 0.03^{\circ}$ and
DEC$=-59.43^{\circ} \pm 0.03^{\circ}$ and it is fully consistent
with the pulsar location. Figure \ref{ProfileSNmap} shows the
source radial profile along the NW-SE axis in the significance
map, together with the profile expected from a point source. The
source appears to be slightly extended and its orientation
coincides with the morphology known at other wavelengths.

\begin{figure}[!h]
\epsscale{1.25} \plotone{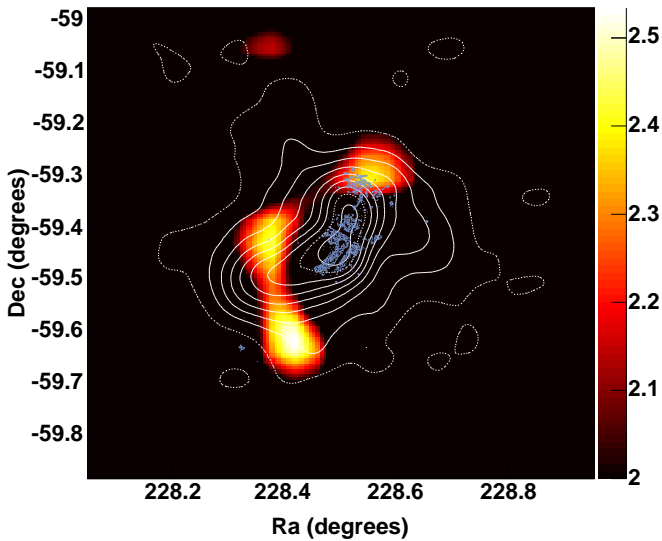}
\caption{Smoothed significance map excess after point source contribution
removal between 17 keV and 40 keV. The white
contours outline the TeV emission as seen by H.E.S.S. and the blue ones correspond to Chandra observation. Two excesses visible in the jet direction (NW-SE) are in global agreement with the H.E.S.S contours baring in mind the smearing due to the PSF. \label{SmoothExcess}}
\end{figure}

With coded mask imaging, sky image pixels are not statistically
independent, but two pixels separated by more than the PSF width
($\sigma_{PSF} = 5.7'$ ) can be considered as independent. After
subtracting the PSF in the significance map (see Figure
\ref{SmoothExcess}) we obtain a confidence level of $4\sigma$ and
$2.3\sigma$ in the low and high-energy bands from the two  residuals
measured along the NW-SE axis.

The source orientation and extension can be evaluated by fitting the intensity images
with a 2D Gaussian with separate widths along each axis and a free
rotation angle. To compare with a true point-source the same fit
has been applied to the XTE J1550 source in the same field of view.

In the low-energy band, XTE J1550 is well fitted by a symmetrical
Gaussian of width $\sigma_{XTE} = 5.85' \pm 0.07'$. It is fully
consistent with the instrument PSF $\sigma_{PSF} = 5.7'$. For PSR
B1509-58, the best fit yields an asymmetric Gaussian with a
rotation angle of $155^{\circ} \pm 4^{\circ}$ from north through
east. Perpendicular to this axis, the source width of $5.90' \pm
0.08'$ is equivalent to that of a point source. Along the NW-SE
direction, the width $\sigma_{NW-SE} = 7.98' \pm 0.07'$ suggests
that the source extends over several arcminutes. The
apparent source width results from the
convolution of the PSF and the true source size. The
latter is found to be $\sigma_{s} = \sqrt{(\sigma_{NW-SE}^{2} -
\sigma_{PSF}^{2})} = 5.53' \pm 0.07'$ along the major axis. In the
energy band from 40 to 100 keV, the source is smaller with
$\sigma_{s} = 3.52' \pm 0.07'$ along the main axis. All the quoted errors
include the statistical uncertainty and a crude estimate of the
impact of the map binning and correlation between adjacent pixels
obtained by moving the gaussian location in the fit.

The fact that the source extension and orientation compare
with the morphology seen in X rays and $\gamma$ rays, as
well as the increased brightness to the southeast, strongly
suggest that the unpulsed emission is dominated by the wind
nebula.

\subsection{Spectroscopy : a possible spectral break near $160$ keV}

\begin{figure}[!h]
\epsscale{1.2}
\plotone{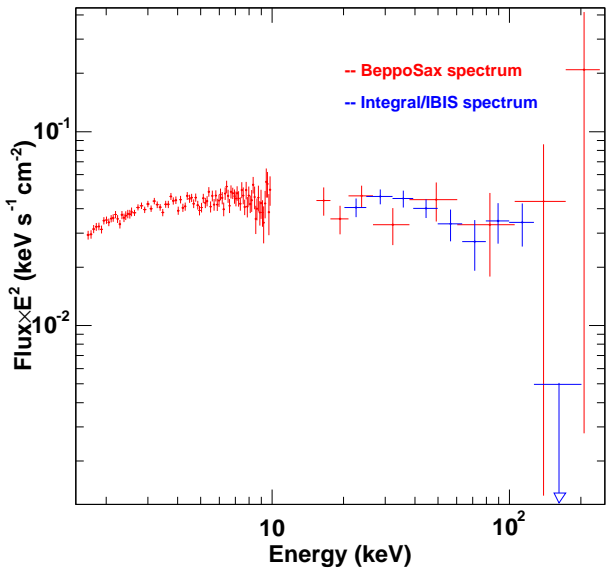}
\caption{Energy flux spectrum of unpulsed emission from the ISGRI and BeppoSAX
data. All error bars and upper limits are given at the 1 $\sigma$
confidence level. \label{SpecBepIntHess}}
\end{figure}

The dilution correction factors due to the source extent with
respect to a point source flux are respectively $1.36$ and $1.19$
in the low and high energy bands. Values at intermediate energies
have been interpolated. The spectrum of the off-pulse emission has
been measured from $17$ keV to $200$ keV using the standard
spectral analysis, then divided by the selected phase width, and
corrected for dilution as described above. It is presented in
Figure \ref{SpecBepIntHess} together with the former BeppoSAX
measurements \citep{Mineo2001}. The INTEGRAL spectrum is well
fitted by an $E^{-2.12 \pm 0.05}$ power law in agreement with the
BeppoSAX  and RXTE results \citep{Mardsen1997}. The total flux from 17 to 200 keV
 amounts to 0.30 $\pm$ 0.02 keV cm$^{-2}$ s$^{-1}$. The upper limit found in
the 130-200 keV bin suggests a spectral break near 160
keV. This upper limit indeed deviates at the 2.9 $\sigma$
level from the extrapolation of the power-law spectrum in
this energy interval. The break is not visible in the
pulsed emission and so cannot be attributed to the pulsar.
It is consistent with the Beppo-SAX measurements at high energy.

\subsection{Discussion and conclusions}

The extent and orientation of the unpulsed emission seen by
INTEGRAL/IBIS nicely match the morphology of the wind
nebula and jet seen at keV energies with ROSAT, Chandra,
and BeppoSAX, and at TeV energies with H.E.S.S. The source width
is not consistent with pure DC emission from the pulsar itself,
 although a large contribution cannot be ruled out. 
There is a close agreement between BeppoSAX and INTEGRAL in spectra
an source elongation. All circumstantial evidence suggests that the measured hard
X-ray emission is dominated by jet emission. The
elongated source appears to be slightly brighter to the south,
with a flux ratio of $\sim 1.4$ between the SE and NW sides.
Larger contrasts are found in X rays ($\sim 5$) and in TeV
$\gamma$ rays ($\sim 3.5$), but the difference may be due
to the poorer angular resolution of ISGRI which cannot separate
the central wind components from the larger scale jet. In the
following discussion we will explore the consequences of a common
origin of the 0.1-160 keV and TeV emissions in the
collimated wind flow or jet along the NW-SE axis.

\begin{figure}[!h]
\epsscale{1.2}
\plotone{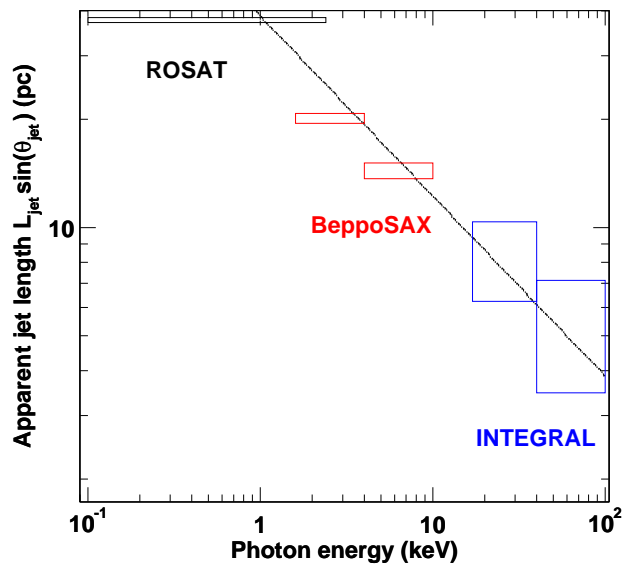}
\caption{Estimates of the jet apparent length, $L_{jet}
sin(\theta_{jet})$, as a function of energy. \label{SyncCool}}
\end{figure}

The relative energy fluxes recorded in X rays and TeV $\gamma$
rays indicate that synchrotron losses dominate over inverse
Compton ones \citep{Hess2005}. Figure \ref{SyncCool} also suggests 
possible decrease of the jet apparent
length, as seen by ROSAT \citep{Trussoni1996}, BeppoSAX
\citep{Mineo2001}, and now IBIS. In a simple
scenario with a linear jet with constant cross-section and
velocity, thus with uniform magnetic field and density to conserve
the magnetic and momentum fluxes, the maximum distance $L_{jet}$
reached by the flow before the pairs burn off their energy scales
as $L_{jet} \propto E_{\gamma}^{-1/2}$ \citep{Pacholczyk1970}:

\begin{equation}
L_{jet} = (1.69 \times 10^{4}) ~\beta_{jet} sin(\theta_{jet}) (\frac{B}{1 \mu
G})^{-3/2} (\frac{E_{\gamma}}{1 keV})^{-1/2} ~pc
\end{equation}

Figure \ref{SyncCool} shows that the data are consistent with this
simple dependence and that radiative aging is effective above keV
energies. This aging is also consistent with the spectral
turn-over seen in Figure
\ref{SpecBepIntHess}. The best $L_{jet} \propto E_{\gamma}^{-1/2}$ fit to
the data in Figure \ref{SyncCool} gives a measure of the product
$\beta_{jet} sin(\theta_{jet}) B^{-3/2}$. The jet outflow has a
typical velocity of 0.5c \citep{Delaney2006} equivalent to the
speed observed in several other wind nebulae (Crab Nebula (0.4c),
Vela (0.3c-0.7c), G11.2-0.3 (0.8-1.4c)). It is consistent with MHD
simulations of relativistic jets (0.5c). A lower limit of 0.3c is
inferred from the Doppler boosting constraint $\beta_{jet}
cos{\zeta} \sim 0.28$. For $\beta_{jet}$ values of 0.3 and 0.5,
we find field strengths of 22 and 33 $\mu G$, respectively. The
large uncertainty in the pulsar distance implies an additional $27
\%$ uncertainty in B. The spectral cut-off possibly seen at 160 keV
then yields maximum electron energies in the jet of 400 and 730
TeV, respectively.

\begin{equation}
E_{e,max} = 230  ~(\frac{E_{\gamma}}{1 keV}) (\frac{B}{1 \mu G})^{-1/2} ~TeV
\end{equation}

The field and maximum energy estimates given above would not apply
if the high-energy part of the INTEGRAL source is dominated by the
toroidal part of the wind instead of the jet as assumed above,
because the rapid increase of the field strength to equipartition
values in the post-shock flow would cool the particles more
efficiently and the INTEGRAL data would not relate to the larger
scale ROSAT and HESS observations.

Following a leptonic scenario and fitting both synchrotron and
inverse Compton radiation, the HESS data yield a mean field of
$17\mu G$ \citep{Hess2005} in reasonable agreement with the above range. Increasing
B to the above values would result in 2 or 3 times fewer particles
in the modelled flow, but the inverse Compton TeV brightness could
still be accomodated owing to the large uncertainty in the actual
infra-red radiation field in the jet vicinity.

In summary, the comparison of the morphology of the wind nebula
above $20$ keV with lower-energy data, and the observation of a possible cut-off near
160 keV are consistent with a simple jet scenario with efficient
synchrotron cooling in a mean magnetic field of order $22-33 \mu
G$ and outflow velocities of 0.3-0.5c. The presence of electrons
with still extreme energies of 400-730 TeV at several parsecs from
the pulsar places a useful constraint on the unknown jet
acceleration process.

\acknowledgements

We gratefully thank the Australian Pulsar Timing
Archive for making all the pulsar epehemerides available to us.

\end{document}